\documentclass{article}

\usepackage{arxiv}

\usepackage[utf8]{inputenc} 
\usepackage[T1]{fontenc}    
\usepackage{hyperref}       
\usepackage{url}            
\usepackage{booktabs}       
\usepackage{amsfonts}       
\usepackage{nicefrac}       
\usepackage{microtype}      
\usepackage{lipsum}
\usepackage{graphicx}
\usepackage{color}
\graphicspath{ {./images/} }
\usepackage{wrapfig}

\usepackage{graphicx}
\usepackage{amsmath}
\usepackage{amssymb}
\usepackage{booktabs}
\usepackage{array}
\usepackage{threeparttable}
\usepackage{epstopdf}
\usepackage{stfloats}
\usepackage{filecontents}
\title{Parameter-Efficient Transformer with Hybrid Axial-Attention for Medical Image Segmentation}

\author{
 Yiyue Hu, Lei Zhang, Nan Mu* \\
  School of Computer Science\\
  Sichuan Normal University\\
  Chengdu 610101\\
  \texttt{nanmu@sicnu.edu.cn} \\
  \And  
   \And
  Lei Liu \\
  School of Science and Engineering\\
  The Chinese University of Hong Kong, Shenzhen\\
  Shenzhen 518172\\
  \texttt{leiliu@link.cuhk.edu.cn} \\
}

\begin{document}
\maketitle
\begin{abstract}
Transformers have achieved remarkable success in medical image analysis owing to their powerful capability to use flexible self-attention mechanism. However, due to lacking intrinsic inductive bias in modeling visual structural information, they generally require a large-scale pre-training schedule, limiting the clinical applications over expensive small-scale medical data. To this end, we propose a parameter-efficient transformer to explore intrinsic inductive bias via position information for medical image segmentation. Specifically, we empirically investigate how different position encoding strategies affect the prediction quality of the region of interest (ROI), and observe that ROIs are sensitive to the position encoding strategies. Motivated by this, we present a novel Hybrid Axial-Attention (HAA), a form of position self-attention that can be equipped with spatial pixel-wise information and relative position information as inductive bias. Moreover, we introduce a gating mechanism to alleviate the burden of training schedule, resulting in efficient feature selection over small-scale datasets. Experiments on the BraTS and Covid19 datasets prove the superiority of our method over the baseline and previous works. Internal workflow visualization with interpretability is conducted to better validate our success.
\end{abstract}


\section{Introduction}
Transformer architectures have become the new paradigm of visual learning tasks, which can capture global spatial relationships on pixels. Benefiting from the pre-training schedule over large-scale datasets \cite{9VIT,25survey}, more inductive biases can be injected into Transformer for downstream tasks. However, when in data-scarce regimes, \textit{e.g.}, medical image analysis, such architecture may suffer from insufficient inductive biases due to lacking intrinsic power of modeling spatial pixel-level information \cite{4MedT}.


The main pathway to capture inductive biases for self-attention was to model the global dependency among elements at different positions in the input sequence. Thus, appropriate position encoding strategies were crucial for the learning capability, since the self-attention mechanism corresponds to a family of permutation equivalence functions \cite{76Explicit}. There were two main attempts to encode position information for self-attention: absolute and relative position encoding. The absolute position encoding (APE) was added to the fixed-length sequence in the input layer to provide fixed position information \cite{1Attention}. To be more flexible, absolute position information can be obtained via trainable parameters \cite{2Bert}. The main disadvantage was that the performance suffers from restricted training dynamics by adding position encoding directly to input tokens forcing their gradients to be the same. Differently, to explore the pairwise relationships among input tokens, relative position encoding (RPE) adopted relative distances in the sequence as position information via the interaction with queries and keys \cite{6}. RPE allowed the self-attention module to capture longer dependencies among tokens. However, RPE required more parameters to calculate position information for both queries and keys, which influences the model complexity and inference efficiency \cite{3AXial-DeepLab,5Axial-Attention}. 

\begin{wrapfigure}{L}{7.5cm}
  \centering
  \includegraphics[width=0.7\linewidth]{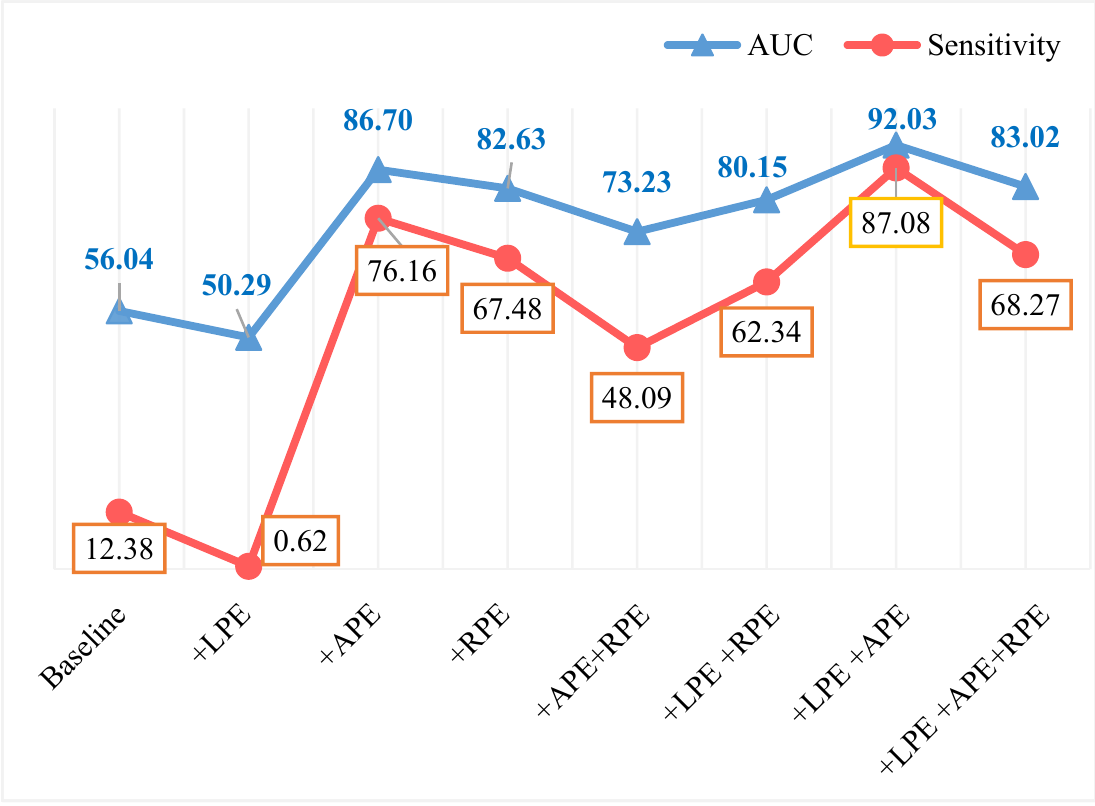}
   \caption{Comparisons of different position encodings, including baseline (without position encoding), absolute position encoding (APE) with fixed frequencies, learnable position encoding (LPE) with flexible frequencies, and relative position encoding (RPE) with learnable position biases. Note that the combination of APE and LPE can obtain higher performance.}
   \label{fig:1}
\end{wrapfigure}

Yet, existing studies of position encoding indicated two surprisingly different phenomenons: (1) \cite{6,8Rethinking,75} proved that RPE can work equally well as APE on natural language processing and computer vision tasks; (2) \cite{9VIT,76Explicit} pointed out that RPE cannot bring apparent gain compared to APE on the classification task. Inspired by this, we took the initiatives to empirically explore the effects of these position encoding strategies, and the main comparisons are shown in Figure \ref{fig:1}. As one of our key contributions, we demonstrate that: (1) APE focuses on the spatial relations and pixel-wise information, which is irreplaceable and crucial for pixel-level image segmentation tasks; (2) RPE emphasizes local patches similarity in shallow layers \cite{8Rethinking}, thus obviously weaker than APE with pixel-level relations. (3) The combination of APE and LPE performs better on small-scale datasets with flexible positional relations.

Motivated by the above, we propose a parameter-efficient transformer with a novel Hybrid Axial-Attention (HAA) to learn more effective pixel-level information. In particular, to alleviate the over-fitting issue on small-scale datasets, our method exploits the combination of APE and LPE to calculate position embeddings, which can be equipped with spatial pixel-wise information and relative position information as inductive bias. To further handle the costly burden of the training schedule, we additionally introduce a gate mechanism to efficiently select adaptive features for the specific task. \textbf{The contributions are as follows: }\textbf{(1)} We develop a parameter-efficient transformer equipped with HAA for medical image segmentation, which can introduce more intrinsic inductive biases for lower computational costs. \textbf{(2)} We empirically investigate the effect of different position encoding strategies and demonstrate that the combination of APE and LPE can perform better on small-scale datasets due to flexible inductive bias. \textbf{(3)} Experiments on BraTS and Covid19 datasets verify the superiority of our method over previous state-of-the-art (SOTA) methods. Internal workflow visualization with interpretability is conducted to better validate the clinical application prospects of our method.


\section{Related Work}
\label{sec:headings}

\subsection{Self-attention Mechanism}
Self-attention plays a fundamental role in Transformer \cite{1Attention,2Bert},  which is capable of modeling the relationship of tokens in a sequence by adding the fixed position encodings by sinusoidal functions or LPE into the input embedding.


\paragraph{Position Encoding.} Position encoding is used to distinguish the different positions in the input sequence, mainly involving APE and RPE strategies \cite{1Attention,68}. The APE emphasizes the absolute position of input tokens from the beginning (one) to the maximum length (sequence length) \cite{8Rethinking}. Absolute positions were combined with the input token embeddings to compute self-attention. For example, the vanilla Transformer \cite{1Attention} adopted sine and cosine functions of different frequencies to encode the absolute position information. To be more flexible, learnable position embedding was proposed to encode absolute position information in \cite{2Bert,9VIT}. In addition, unlike the position encoding used in previous work \cite{1Attention,2Bert,9VIT,68} which were predefined and input-agnostic, a new encoding called CVPT was generated on-the-fly and conditioned on the local neighborhood of an input token. The RPE focuses on the relative distance between input elements and learns the pairwise relation of tokens. For instance, Transformer-XL \cite{70Transformer-XL} enabled learning dependency beyond a fixed length without disrupting temporal coherence by introducing the sinusoid formulation for relative position encoding. \cite{6} introduced a learnable vector to represent two arbitrary positions to efficiently extend the self-attention mechanism. To better capture long-range interaction with precise position information, \cite{3AXial-DeepLab} proposed to utilize the RPEs for all query, key, and value tokens.

\paragraph{Axial-attention.} To reduce computation complexity and allow performing attention within a larger or even global region, the 2D self-attention is factorized into two 1D self-attention, called Axial-attention. \cite{5Axial-Attention} presented an axial transformer using axial self-attention layers and a shift operation to efficiently build full receptive fields of multidimensional tensors. To get the affinities with global position information, a position bias term for RPEs was used in axial attention to exploring position-sensitive information \cite{6}. Furthermore, a position-sensitive method proposed a qkv-dependent position bias \cite{3AXial-DeepLab} to capture precise position information at a reasonable computation overhead. To alleviate overfitting on the small-scale datasets, MedT \cite{4MedT} proposed an additional control axial-attention block to remit the inaccuracy of relative position bias, which can achieve better performance with lower computational complexity.

\subsection{Medical Image Segmentation}
Recently, various Transformer variants have achieved impressive performance in medical image processing domain \cite{25survey,26MedicalSuervey}. Typically, to maintain the advantages of Transformers and U-Net \cite{12UNet}, TransUNet \cite{12UNet} employed a hybrid CNN-Transformer architecture to leverage both detailed high-resolution spatial information from CNN features and the global context encoded by Transformers. Similarly, combining self-attention and convolution in parallel style, TransFuse \cite{59TransFuse} proposed a novel fusion technique BiFusion module to fuse the multi-level features. Swin-UNet \cite{55SwinUnet} presented a pure Transformer-based U-shaped encoder-decoder network outperforming fully convolution or the combination of transformer and convolution methods. For MRI Brain Tumor segmentation, TransBTS \cite{73TransBTS} exploited 3D CNN with transformer to extract the volumetric spatial features and progressively upsampling strategy.

In this work, we mainly focused on Magnetic Resonance Imaging (MRI) and Computed Tomography Angiography (CTA) image segmentation which plays a fundamental role in various medical evaluations \cite{28eg}. Furthermore, MRI, CTA, and other medical images are frequently destroyed by the high complexity of noise, signal dropout, illegible structures, and poor contrast along tissue boundaries \cite{29Deep}. Furthermore, the existing automatic segmentation method generally lacks interpretability \cite{27interpretability} for clinical applications. 

\section{The Proposed Method}

\subsection{Preliminary}

\paragraph{Self-attention Mechanism.} Given an input feature map $ X\in{R^{C_\text{in} \times H \times W}} $ with channels $C_\text{in}$, height $H$, and width $W$, the query $Q$, key $K$ and value $V$ representations are obtained via different projections by:
\begin{equation}
    Q = W_qX, \quad K = W_kX, \quad V = W_vX,
    \label{eq:1}
\end{equation}
where $W_q, W_k, W_v$ are the parameter matrices. Then the weighted sum output $Y \in{R^{C_\text{out} \times H \times W}}$ over value representations is calculated by a self-attention layer using the following formulation:
\begin{equation}
    y_{i j}=\sum_{h=1}^{H} \sum_{w=1}^{W} \operatorname{softmax}\left(q_{i j}^{T} k_{h w}\right) v_{h w},
    \label{eq:2}
\end{equation}
where $q_{ij} ($or $k_{ij}, v_{ij})$ denotes query (or key, value) vector at location $i\in[1, H]$ and $j\in[1, W]$. $\operatorname{softmax}(\cdot)$ denotes the normalized similarity score over the pairwise similarity matrix between each location of $Q$ and $K$, which takes $ O((HW)^2)$ computational complexity.

\paragraph{Axial-attention Mechanism.} To reduce the computational costs of self-attention, axial-attention mechanism is proposed to enable building a fully self-attention model. Concretely, there are two types of axial-attention: height-axis and width-axis. The height-axis type is formulated as:
\begin{equation}
    y_{i j}=\sum_{h=1}^{H} \operatorname{softmax}\left(q_{i j}^{T} k_{h j}\right) v_{h j},
    \label{eq:3}
\end{equation}
where the computation cost is $O(HHW)$ to compute the pairwise similarities between each location of the height-axis feature vector. Similarly, the width-axis axial-attention is along the width axis with the computation cost $O(HWW)$. In this work, axial-attention is the default attention mechanism for lower computational complexity.

\subsection{Position Encoding}
Both self-attention and axial-attention operations in Transformer are permutation-invariant, \textit{i.e.}, the position information about the order of tokens or pixel positions is ignored. Thus, various position encoding strategies are proposed to encode such crucial prior information.

\paragraph{Absolute Position Encoding (APE).} The vanilla self-attention incorporates sequence information into query, key, and value representations via absolute positions \cite{1Attention}:
\begin{equation}
    x_{i} = x_{i} + p_{i},
    \label{eq:4}
\end{equation}
where $x_i$ is $i$-th row (or column) vector in $X$ for height-axis (or width-axis) attention. $p_i$ is absolute position vector at position $i$, which is calculated by:

\begin{equation}
    \left\{\begin{array}{ll}
    p_{i, 2 t} & =\sin \left(i / 10000^{2 t / d}\right) \\
    p_{i, 2 t+1} & =\cos \left(i / 10000^{2 t / d}\right),
    \end{array}\right.
    \label{eq:5}
\end{equation}
where $p_{i,2t}$ is the 2t-$th$ element of $p_i$. The absolute position encoding can reflect spatial location information among pixels using sinusoidal encoding. One variant of APE mainly obtains $p_i$ position encoding via more flexible frequencies, we call it LPE for short \cite{2Bert}.

\textbf{Relative Position Encoding (RPE).} Considering the pairwise distance between sequence elements in the attention layer, \cite{6} added a trainable position vector to the input $x_i$ to encode relative distances:
\begin{equation}
   y_{i j}=\operatorname{softmax}(W_{q}x_{i} (W_{k}x_{j} +p_{i j}^{k})\left(W_{v}x_{j} +p_{i j}^{v}\right)),
\label{eq:6}
\end{equation}
where $ p_{i j}^{k}$ and $ p_{i j}^{v} $ are trainable weights of RPE on values and keys, respectively. Based on the hypothesis that precise information is not useful beyond a certain distance, a clip function is denoted as $ clip(x,k)=max(-k, min(k,x))$ and is applied on the $ p_{i j}^{k}, p_{i j}^{v} $.


\paragraph{Combination of APE and RPE.} In general, APE and RPE can be combined via the following formulation:
\begin{equation}
\begin{aligned}
   x_i&=x_i+p_i, \quad x_j=x_j+p_j,\\
    y_{i j}&=\operatorname{softmax}(q_i k_j+q_i r^q +k_j r^k)\left(v_j+r^v\right),
    \label{eq:7}  
\end{aligned}
\end{equation}
where $q_i=W_qx_i, k_j=W_kx_j, v_j=W_vx_j$, and $r^q, r^k, r^v$ are trainable vectors called position bias terms.

\subsection{Empirical Study}

To clarify the influence of position encoding on small-scale datasets, we conducted an empirical study on the BraTS dataset (900 training samples) to explore the effect of axial-attention with different position encoding strategies: (1) baseline without position encoding; (2) LPE, \textit{i.e.}, APE with learnable encoding; (3) APE with fixed sinusoidal encoding; 4) RPE with a trainable position bias term; 5) their combinations. Quite noticeably, the combination of LPE and APE obtains SOTA results over all protocols, especially on AUC and Sensitivity. Detailed analysis is as follows:


\paragraph{Quantitative Analysis.} As shown in Table \ref{tab:1}, positional information is necessary for the segmentation task to capture sequential correlations of different positions. For the perspective of absolute position and relative distance, APE and RPE can embed spatial pixel-wise information and relative distance information into attention calculation. However, the combination of APE and RPE may hurt model performance, since the trainable position bias term in RPE may affect the properties of APE, \textit{e.g.,} Monotonicity and Symmetry \cite{83OP},  
which easily leads to confused contextual information. Similarly, the combination of LPE and RPE also obtains relatively worse performance. Besides, LPE can adaptively allocate different frequencies in a data-driven way, \textit{i.e.}, not only limited to fixed patterns, however, it is noticed that only LPE performs worse than APE and RPE over all evaluation protocols, since its learnable positional parameters require large-scale training data to capture reliable positional information. Overall, APE with fixed frequencies (\textit{e.g.}, $i/10000^{2 t / d}$) and LPE with flexible frequencies can benefit from each other to obtain higher performance.

\begin{table}[!t]
  \caption{Quantitative comparison of various position encodings.}
  \centering
    \begin{tabular}{lccccc}
    \toprule
    \textbf{Methods} & \multicolumn{1}{c}{\textbf{AUC}} & \multicolumn{1}{c}{\textbf{WF}} & \multicolumn{1}{c}{\textbf{Dice}} & \multicolumn{1}{c}{\textbf{Jacc}} & \multicolumn{1}{c}{\textbf{Sensi}} \\
    \midrule
    Baseline & 56.04 & 18.18 & 16.24 & 11.51 & 12.38 \\
    +LPE  & 50.29 & 1.17  & 1.02  & 0.62  & 0.62 \\
    +APE  & 86.70  & 75.27 & 75.00    & 64.54 & 76.16 \\
    +RPE  & 82.63 & 77.764 & 74.64 & 64.13 & 67.48 \\
    +APE+RPE & 73.23 & 64.91 & 59.46 & 46.80  & 48.09 \\
    +LPE+RPE & 80.15 & 72.85  & 69.80  & 58.28 & 62.34 \\
    +LPE+APE & \textcolor[rgb]{ .753,  0,  0}{\textbf{92.03}} & \textcolor[rgb]{ .753,  0,  0}{\textbf{78.81}} & \textcolor[rgb]{ .753,  0,  0}{\textbf{80.20}} & \textcolor[rgb]{ .753,  0,  0}{\textbf{69.81}} & \textcolor[rgb]{ .753,  0,  0}{\textbf{87.08}} \\
    \bottomrule
    \end{tabular}%
  \label{tab:1}%
\end{table}%

\begin{figure}[t]
  \centering
  \includegraphics[width=1\linewidth]{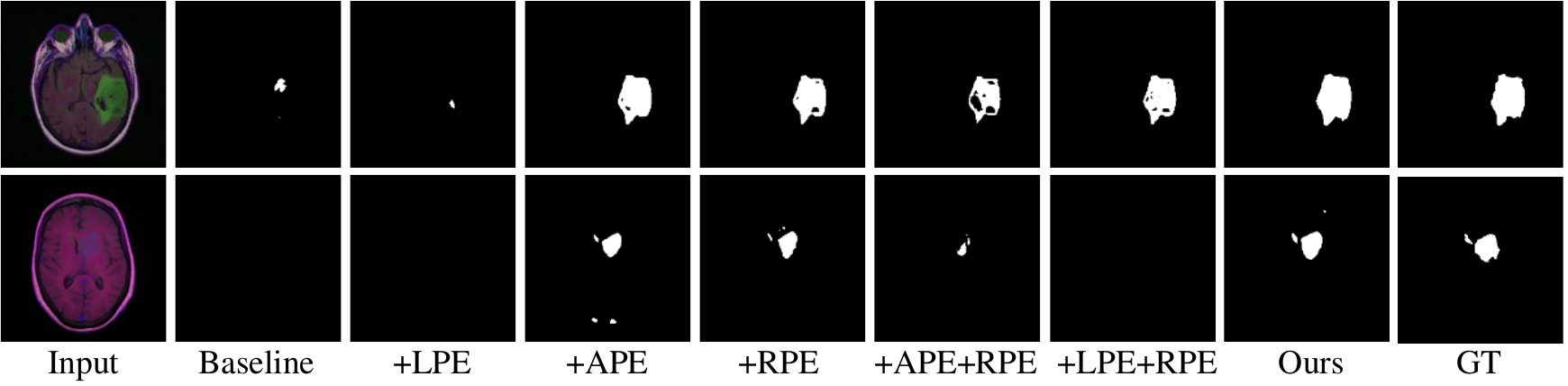}
   \caption{Qualitative results of different position encoding strategies on BraTS dataset.}
   \label{fig:2}
\end{figure}

\paragraph{Qualitative Analysis.} We visualized the typical results shown in Figure \ref{fig:2}. It can be observed that LPE cannot locate the lesion region due to a lack of inductive bias under small-scale medical datasets, it is hard to optimize the learnable parameters, similarly for baseline. Note that since RPE encodes relative distance via the interaction with queries and keys, when the learnable bias is inaccurate enough would cause a loss of spatial position information between pixels. Furthermore, the integration of RPE and APE/LPE is unable to capture the complete lesion regions since it lacks reliable contextual information. However, APE with fixed frequencies can be used as prior knowledge to encode each pixel on the image to carry spatial sequence information (can be understood as a generalization of one-dimensional sequences), making the feature tensor with spatial position information, and APE with flexible frequencies can correct prior errors brought by APE.

Taking the above discussions into account, we draw the following conclusion: APE with sinusoidal frequencies and LPE with flexible frequencies can import inductive bias to encode spatial sequence information and relative position information well, and obtain better performance.

\subsection{Parameter-Efficent Network}

\begin{figure*}[t]
  \centering
  \includegraphics[width=1\linewidth]{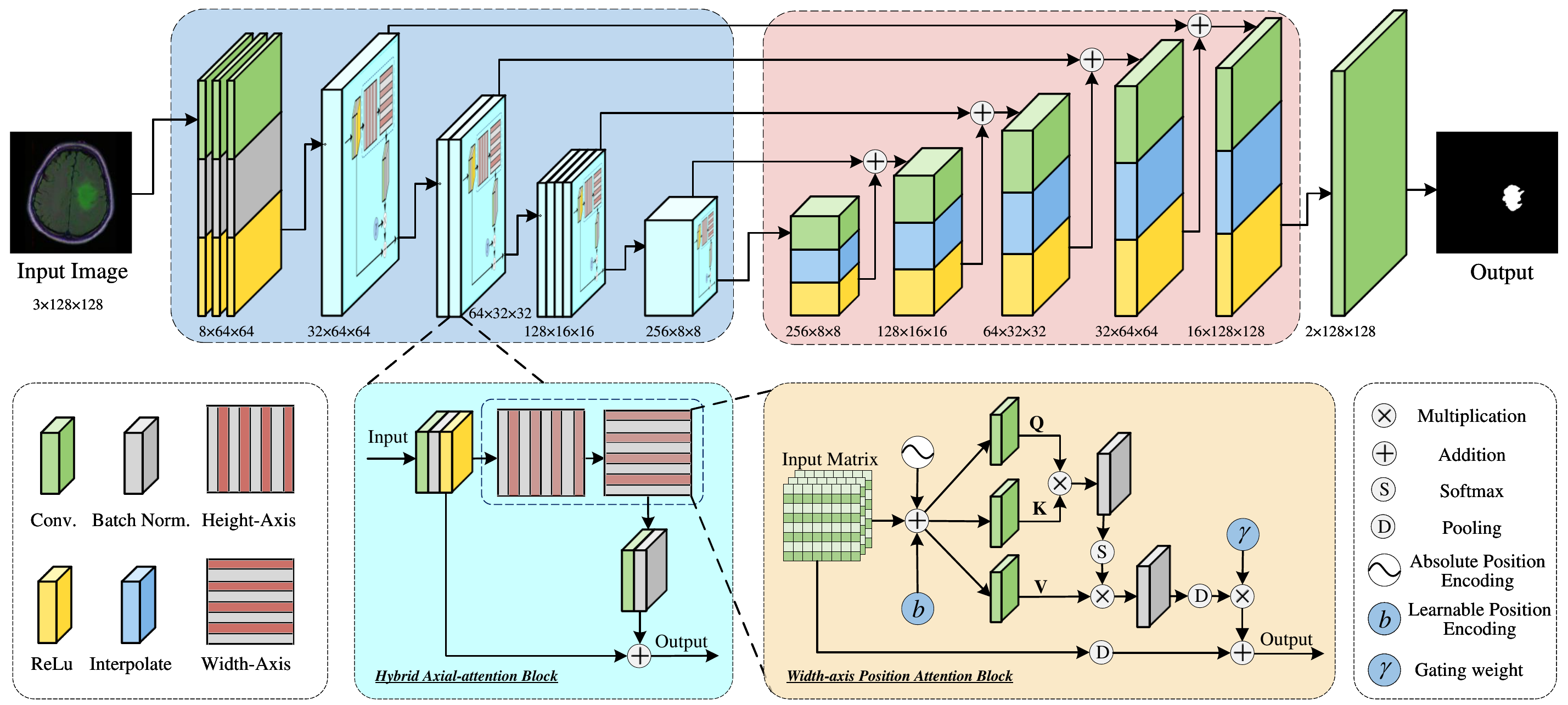}
   \caption{The architecture of the proposed model. The hybrid axial-attention block is the basic building block of the encoder, which propagates information along height-axis and width-axis sequentially to model the long-range dependency. A width-axis position attention block shows the whole process of calculating the attention score through the integration of LPE and APE.}
   \label{fig:3}
   \vspace{-0.4cm}  
\end{figure*}

\paragraph{Overall Architecture.}
As shown in Figure \ref{fig:3}, the overall network adopted an encoder-decoder architecture following the design of ResUNet \cite{24ResUNet}. In the encoder stage, a convolutional block was utilized to extract high-resolution with semantically weak features \cite{12UNet}. Ultimately, we designed the HAA-based encoder layers to model the global-range dependency with lower computational cost; In the decoder stage, the decoder layers were exploited to assemble a more precise output by combing high-resolution features from the contracting path with upsampled output.

\paragraph{Notations.} Let $X \in \mathbb{R}^{C_\text{in} \times H \times W}$, $F^c \in \mathbb{R}^{C_\text{out}  \times H  \times W}$ denote the input and output feature maps respectively, where $H, W$ indicate the height and width. $X_{ij} \in{\mathbb{R}^{C_\text{in}}} $ ($F^c_{ij}  \in{\mathbb{R}^{C_\text{out}}}$) denotes pixel at the position $(i,j)$ of $X$ ($F^c$).

\subsubsection{Hybrid Axial-attention (HAA) based Encoder}

\paragraph{Convolutional Block.} For the input feature map $X$, we used a general convolution operation $\mathcal{C}$ to extract local semantic information via stacking layers with different convolutional kernels $\mathcal{K}$, \textit{i.e.}, $\mathcal{K} \in{\mathbb{R}^{C_\text{in}  \times C_\text{out}  \times k  \times s \times p}}$ \textit{i.e.}, where $C_\text{in}$ and $C_\text{out}$ are the input and output channel size, $k$ is the kernel size, $s$ is the stride and $p$ is the padding. We exploited the form of Conv-BN-ReLu to build a convolutional layer:
\begin{equation}
    F^{c}= \text{ReLu}(\text{BN}(\mathcal{C}(X, \mathcal{K}))),
    \label{eq:8}
\end{equation}
where ReLu is the activation function and BN is the batch normalization operation \cite{14BN}.

\paragraph{Adaptive Position Embedding.}
Motivated by the empirical study, we presented to utilize the elegant integration of LPE and APE called adaptive position embedding (ADPE), which can be equipped with spatial pixel-wise information and relative position information as the inductive bias for better representations. Concretely, given the feature map $F_c$ outputted by the convolutional block, the feature map $\hat{F}^{c}$ with positional information is obtained by:
\begin{equation}
    \hat{F}^{c} = F^{c} \oplus P^l \oplus P^s,
    \label{eq:9}
\end{equation}
where $\oplus$ denotes the element-wise addition, $P^l$ is the learnable absolute position matrix, and $P^s$ is the absolute position matrix ($P^s_{ij}$ in obtained by Eq. \ref{eq:5}).



\paragraph{Hybrid Axial-attention.} 
Based on the convolutional block, the hybrid axial-attention mechanism is to capture global dependency information, which combines axial-attention along the height-axis and width-axis directions in the vector space \cite{5Axial-Attention}. \textbf{(1) Height-axis Attention.} Given the feature map $\hat{F}_C$ with position information,
$\hat{F}_C$ is firstly transformed into $\mathbb{R}^{W \times C \times H}$ based on permute operation. Then height-axis attention can reduce the complexity from $O(HHWW)$ to $O(HHW)$, which calculates the unidirectional affinities along height direction as follows:
\begin{equation}
\begin{aligned}
    &Q=W_q\hat{F}^c, \quad K=W_k\hat{F}^c, \quad V=W_v\hat{F}^c, \\
    &F^{h}_{ij} = \sum_{h=1}^{H} \text{Avg} (\text{BN}((\operatorname{softmax}\left(q_{i j}^{T} k_{h j}\right) v_{h j})),
    \label{eq:10}
\end{aligned}
\end{equation}
where $\text{Avg}$ is a global average pooling operation used to reduce the number of parameters and obtain lower resolution feature tensors, and $F^{h}$ is the attention-based output along the height axis. $q_{ij}, k_{ij}, v_{ij}$ denotes the value at position $(i, j)$ of $Q, K, V$. 
\textbf{(2) Width-axis Attention.} The output of height-axis attention typically captures the local dependencies along height-axis, which cannot aggregate the global information, even if it is stacked several times. Thus we considered sequentially stacking height-axis and width-axis attention to propagate information along two directions to better obtain the global-range dependencies. More specifically, we permuted $F^{h}$ into $\mathbb{R}^{W \times C \times H}$, then the width-axis attention was calculated to obtain the output $F^w$ following Eq. \ref{eq:10}, where the main difference is the direction.


\paragraph{Gating Mechanism.}
To alleviate the over-fitting issue on the small-scale datasets, it is
worth exploring more inductive biases into the network via regularization techniques \cite{80}, which can improve the search for solutions and slightly boost performance. Specifically, when the global affinities is activated on large-scale datasets, the gating weight tends to converge to higher values. On the other hand, if the calculated affinities are inaccurate on small-scale datasets, the gating weight tends to approximately converge to zero.  Motivated by this, we introduced gate mechanism to enforce the model focus on the most informative components of features \cite{81}. In detail, the gating weight is applied on both height-axis and width-axis position attention block, which can be expressed as:
\begin{equation}
\begin{aligned}
 &F^h = \gamma_1 \cdot \text{HA}(\hat{F}^c)  + \text{Avg} (F^c), \\
 & F^w = \gamma_2 \cdot \text{WA}(F^h) + \text{Avg} (F^h),
\end{aligned}
\label{eq:11}
\end{equation}
where $\text{HA}, \text{WA}$ are height-axis and width-axis axial-attention layers with different gating weights $\gamma_1$ and $\gamma_2$.

\subsubsection{Decoder}
To increase the resolution of the contracting network, the decoder consists of five successive layers and each layer adopts a Conv-BiL-ReLu architecture, where BiL denotes the Bilinear interpolation operation as follows: 
\begin{equation}
    \hat{O} = \text{ReLu}(\text{Bilinear}(\mathcal{C}(F^w, \mathcal{K}))) \oplus F^w,
    \label{eq:12}
\end{equation}
where $\hat{O}$ is the output with global contextual information. Note that each layer has the same kernel size. A $3\times 3$ convolution is followed by the decoder to get the final prediction results. In particular, a skip connection is exploited after each decoder layer to obtain a more precise prediction. Concretely, before the $i$-th decoder layer, there is a feature fusion operation between the upsampled feature from the previous ($i-1$) decoder layer and a high-resolution feature from the corresponding encoder layer with same resolution.

\subsection{Loss Function}
Binary Cross-entropy (BCE) Loss \cite{79BCELoss} is the most widely used loss in segmentation tasks to measure the difference between the prediction map and the ground truth distribution, which can be defined as:
\begin{equation}
    L=-\frac{1}{n} \sum_{k=1}^{n} G_{k} \log Y_{k}+\left(1-G_{k}\right) \log \left(1-Y_{k}\right),
  \label{eq:18}
\end{equation}
in which $n=H\times W$ denotes the pixel number of the input image, $ G \in {(0,1)} $ is the ground truth and $Y$ is the probability of segmentation lesion region. 


\section{Experiments}

\subsection{Experimental Setup}
\paragraph{Datasets.} In this work, the experiments were evaluated on two datasets. (1) Brain Tumor Segmentation (BraTS) dataset \cite{60} , which contains brain MRI images together with manual FLAIR abnormal segmentation masks (image of tumor region outlined in white pixels on black background) \cite{29Deep}, involving about 110 different patients from five different institutions diagnosed with LGG (considered lower-grade gliomas), in total of 3929 images. To evaluate the performance of our model more effectively, we washed out most of the images without lesion regions, leaving 1373 images. (2) COVID-19 CT segmentation (Covid19) dataset \cite{covid19}, which consists of 100 axial CTA images from different patients that were collected by the Italian Societ of Medical and International Radiology, a total of 978 images with annotated infection regions were used for our experiments. 

\paragraph{Implementations.}
Our model was implemented in PyTorch framework. Each image was resized to $256 \times 256$. A RTX 3090 GPU was used for both training and testing. The network parameters were initialized by standard normal distribution. Adam optimizer was utilized for optimization. The initial learning rate is 0.003, weight-decay is $1e^{-5}$, and momentum is 0.9. A total of 400 epochs were executed with a batch size of 1. The entire procedure took about 1.2 hours.

\paragraph{Protocols.} To evaluate the performance of the proposed model against the other ten SOTA models, we used random sampling to divide BraTS dataset into 900 images for training, 230 images for valuating, 243 images for testing. The Covid19 dataset was split into 600 images for training, 246 images for valuating, 132 images for testing. Nine segmentation indicators were exploited: 1) AUC score (Area Under Curve) ; 2) MAE score (Mean Absolute Error); 3) WF (Weighted F-measure) score; 4) OR (Overlapping Ratio) score; 5) Dice score; 6) Jaccard score; 7) Sensitivity; 8) Specificity; 9) Parameters (\#M) and FLOPs (\#GMac).

\paragraph{Comparisons.} CNNs-based and Transformer-based models were the mainstay of image segmentation. Therefore, we divided comparison experiments into Convolution-based models and Transformer-based models. UNet \cite{12UNet}, ResUNet \cite{24ResUNet}, SENet \cite{10SENet}, CANet \cite{23CANet}, Inf-Net \cite{20InfNet} and BasNet \cite{22BasNet} are convolution-based models, AxialNet \cite{5Axial-Attention}, Axial-DL \cite{3AXial-DeepLab}, Gate Axail \cite{4MedT}, MedT \cite{4MedT} and ours belong to Transformer-based models.

\subsection{Comparison Results}
\paragraph{Quantitative Results.} The segmentation results of BraTS dataset and Covid19 dataset are shown in Table \ref{tab:2} and Table \ref{tab:3}, respectively. Note that CNNs-based models perform better than Transformer-based models, which are difficult to train on small-scale datasets. The proposed method is able to overcome this issue by exploiting spatial pixel-level information and gate mechanism as the inductive bias. Our method outperforms CNNs-based by approximately 7.60\%, and Transformer-based baselines by 40.10\% over all metrics. Although CNN-based model has a slightly higher score in some metrics, its Params and FLOPs are much larger than ours, even much more than 10 times.

\begin{table*}[h!t]
 \caption{Comparison results of the proposed method with CNNs-based and Transformer-based baselines on BraTS dataset. Red color means the best result while green color indicates the second best result.}
  \centering
    \begin{tabular}{lccccccccc}
    \toprule
    \textbf{Methods} & \multicolumn{1}{c}{\textbf{Params}} & \multicolumn{1}{c}{\textbf{AUC}} & \multicolumn{1}{c}{\textbf{MAE}} & \multicolumn{1}{c}{\textbf{WF}} & \multicolumn{1}{c}{\textbf{OR}} & \multicolumn{1}{c}{\textbf{Dice}} & \multicolumn{1}{c}{\textbf{Jaccard}} & \multicolumn{1}{c}{\textbf{Sensitivity}} & \multicolumn{1}{c}{\textbf{Specificity}} \\
    \midrule
    UNet \cite{12UNet} & 13.40   & 80.77 & 0.91  & 74.69 & 60.61 & 71.40  & 60.59 & 63.67 & 99.88 \\
    ResUNet \cite{24ResUNet} & 11.91 & 83.78 & 0.85  & 75.04 & 63.71 & 73.46 & 63.67 & 69.98 & 99.76 \\
    SENet \cite{10SENet} & 11.98 & 82.35 & 0.88  & 76.39 & 63.12 & 73.61 & 63.09 & 66.86 &  \textcolor[rgb]{ 0,  .69,  .314}{\textbf{99.89}} \\
    CANet \cite{23CANet} & 11.98 & 85.27 & 0.78  & 79.54 & 67.41 & 77.54 & 67.38 & 72.91 & 99.83 \\
    InfNet \cite{20InfNet} & 33.12 & \textcolor[rgb]{ .753,  0,  0}{\textbf{93.29}} & 3.68  & 37.3  & \textcolor[rgb]{ .753,  0,  0}{\textbf{73.2}} & \textcolor[rgb]{ 0,  .69,  .314}{\textbf{80.24}} & 70.18 & \textcolor[rgb]{ .753,  0,  0}{\textbf{94.02}} & 98.02 \\
    BasNet \cite{22BasNet} & 87.06 & 89.30  & 0.91  & \textcolor[rgb]{ 0,  .69,  .314}{\textbf{79.4}} & 71.12 & 79.31 & \textcolor[rgb]{ 0,  .69,  .314}{\textbf{71.11}} & 81.57 & 99.45 \\ \midrule
    AxialNet \cite{5Axial-Attention}& 1.35  & 56.04 & 2.54  & 18.18 & 12.87 & 16.24 & 11.51 & 12.38 & \textcolor[rgb]{ .753,  0,  0}{\textbf{99.99}} \\
    Axial-DL \cite{3AXial-DeepLab} &  \textcolor[rgb]{ 0,  .69,  .314}{\textbf{1.33}}  & 84.86 & 1.02  & 70.06 & 60.17 & 70.29 & 60.13 & 72.47 & 99.51 \\
    Gated Axial \cite{4MedT} &  \textcolor[rgb]{ 0,  .69,  .314}{\textbf{1.33}}  & 85.47 & 1.3   & 68.54 & 58.23 & 69.19 & 58.23 & 73.8  & 99.31 \\
    MedT \cite{4MedT} & 1.56  & 86.11 & \textcolor[rgb]{ 0,  .69,  .314}{\textbf{0.81}} & 77.5  & 66.24 & 76.56 & 66.23 & 74.75 & 99.73 \\ \midrule
    \textbf{Ours} & \textcolor[rgb]{ .753,  0,  0}{\textbf{1.31}} & \textcolor[rgb]{ 0,  .69,  .314}{\textbf{89.79}} & \textcolor[rgb]{ .753,  0,  0}{\textbf{0.70}} & \textcolor[rgb]{ .753,  0,  0}{\textbf{82.02}} & \textcolor[rgb]{ 0,  .69,  .314}{\textbf{71.87}} & \textcolor[rgb]{ .753,  0,  0}{\textbf{81.79}} & \textcolor[rgb]{ .753,  0,  0}{\textbf{71.87}} & \textcolor[rgb]{ 0,  .69,  .314}{\textbf{82.31}} & 99.68 \\
    \bottomrule
    \end{tabular}%
  \label{tab:2}%
\end{table*}%

\begin{table}[!t]
 \caption{Quantitative results on Covid19 dataset. The best result is marked in red, while the second best is in green.}
\setlength{\tabcolsep}{4pt}
  \centering
    \begin{tabular}{lccccccccc}
    \toprule
    \textbf{Methods} & \multicolumn{1}{c}{\textbf{AUC}} & \multicolumn{1}{c}{\textbf{MAE}} & \multicolumn{1}{c}{\textbf{WF}} & \multicolumn{1}{c}{\textbf{Dice}} & \multicolumn{1}{c}{\textbf{Jacc}} & \multicolumn{1}{c}{\textbf{Sensi}} \\
    \midrule
    UNet  & 81.03 & 2.71  & 77.28 & 74.07 & 61.41 & 67.69 \\
    ResUNet  & 86.47 & 1.74  & \textcolor[rgb]{ .753,  0,  0}{\textbf{86.75}} & \textcolor[rgb]{ .753,  0,  0}{\textbf{83.79}} & \textcolor[rgb]{ 0,  .69,  .314}{\textbf{73.60}} & 79.25 \\
    \midrule
    InfNet  & \textcolor[rgb]{ 0,  .69,  .314}{\textbf{91.00}} & 1.81  & 85.99 & 79.75 & 67.10  & \textcolor[rgb]{ 0,  .69,  .314}{\textbf{94.80}} \\
    BasNet  & 85.44 & 2.87  & 73.00    & 71.07 & 57.74 & 79.85 \\
    AxialNet & 89.82 & 9.81  & 46.78 & 55.42 & 40.00    & 94.00 \\
    Axial-DL & 78.00 & 2.57  & 77.86 & 72.38 & 59.24 & 60.83 \\
    Gate Axial  & 81.83 & 2.25  & 81.81 & 77.65 & 65.43 & 69.18 \\
    MedT  & \multicolumn{1}{c}{75.45} & 2.96  & 73.25 & 67.27 & 53.61 & 55.28 \\
    \midrule
    \textbf{Ours}  & \textcolor[rgb]{ .753,  0,  0}{\textbf{92.06}} & \textcolor[rgb]{ .753,  0,  0}{\textbf{1.29}} & \textcolor[rgb]{ 0,  .69,  .314}{\textbf{86.13}} & \textcolor[rgb]{ 0,  .69,  .314}{\textbf{80.45}} & \textcolor[rgb]{ .753,  0,  0}{\textbf{79.75}} & \textcolor[rgb]{ .753,  0,  0}{\textbf{94.88}} \\
    \bottomrule
    \end{tabular}%
  \label{tab:3}%
\end{table}%

\begin{figure}[t]
  \centering
  \includegraphics[width=1\linewidth]{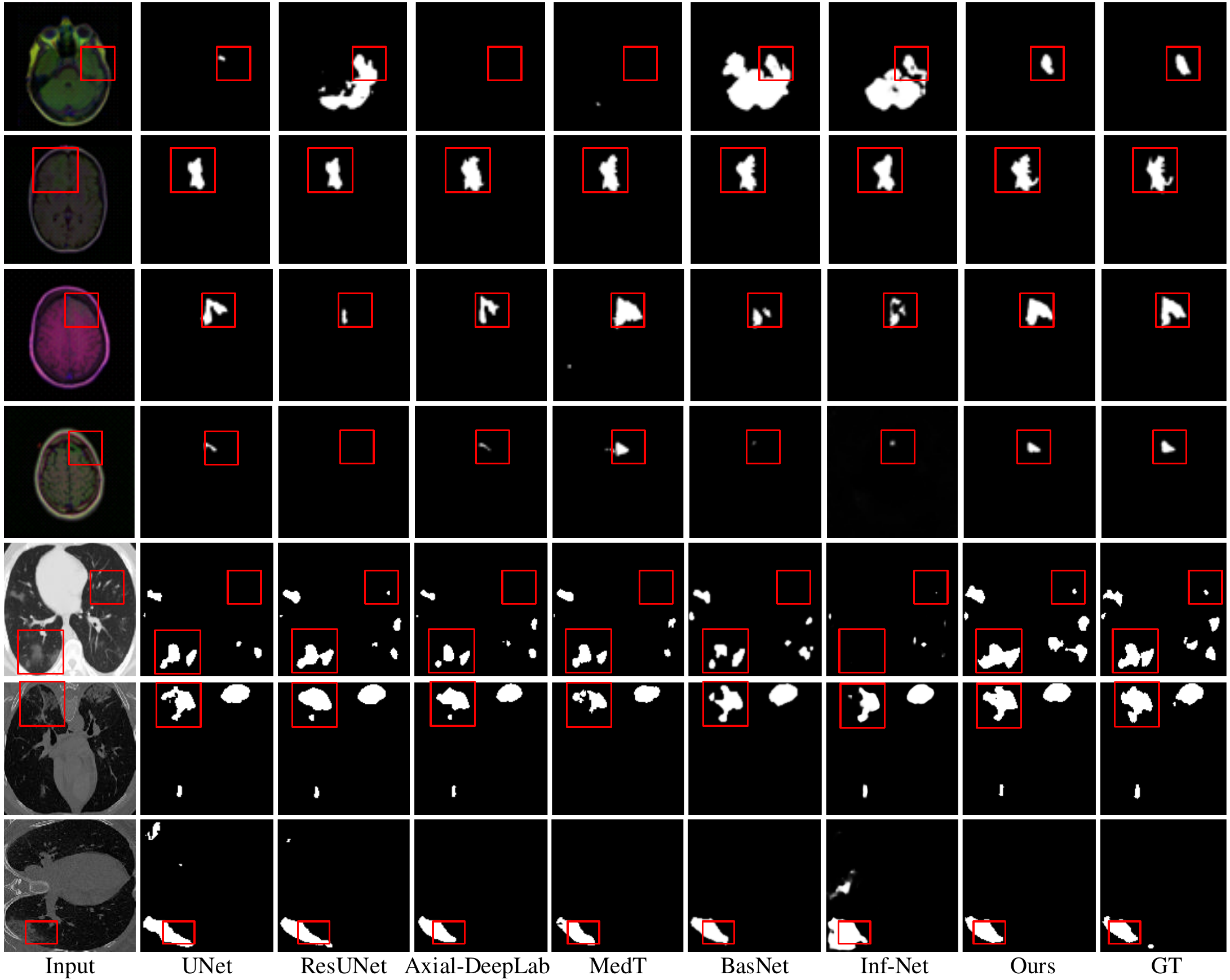}
   \caption{Comparisons of visualized segmentation results on BraTS (row-1 to row-4) and Covid19 (row-5 to row-7) datasets. The red boxes highlight the regions where our method performs exactly better than the others.}
   \label{fig:4}
\end{figure}

\paragraph{Qualitative Results.} The visualized results are shown in Figure \ref{fig:4}. Compared with Transformer-based methods such as Axial-DeepLab and MedT, our method can perform better on the both small (row-4, col-8) and large lesion regions (row-3, col-8) by integrating more flexible position embeddings and gate mechanism, while Axial-DeepLab cannot encode global context information due to worse position biases on small-scale datasets (illustrated in row-1, col-4 and row-4, col-4). 
Besides, RPE-based MedT only captures relatively worse pixel-wise dependencies (see row-1, col-5), although utilizing four gate parameters to alleviate the training burden. Furthermore, CNNs-based methods like BasNet (see col-6) and InfNet (see col-7) lack the ability of modeling long-range dependency since the convolutional kernel attends to only a local-subset of pixels in the whole image, so they are forced to focus on local patterns. In summary, our model can accurately model the long-range dependency while reducing computational complexity, which is very suitable for small-scale medical datasets.

\subsection{Clinical Interpretability}

\begin{figure*}[t]
  \centering
  \includegraphics[width=1\linewidth]{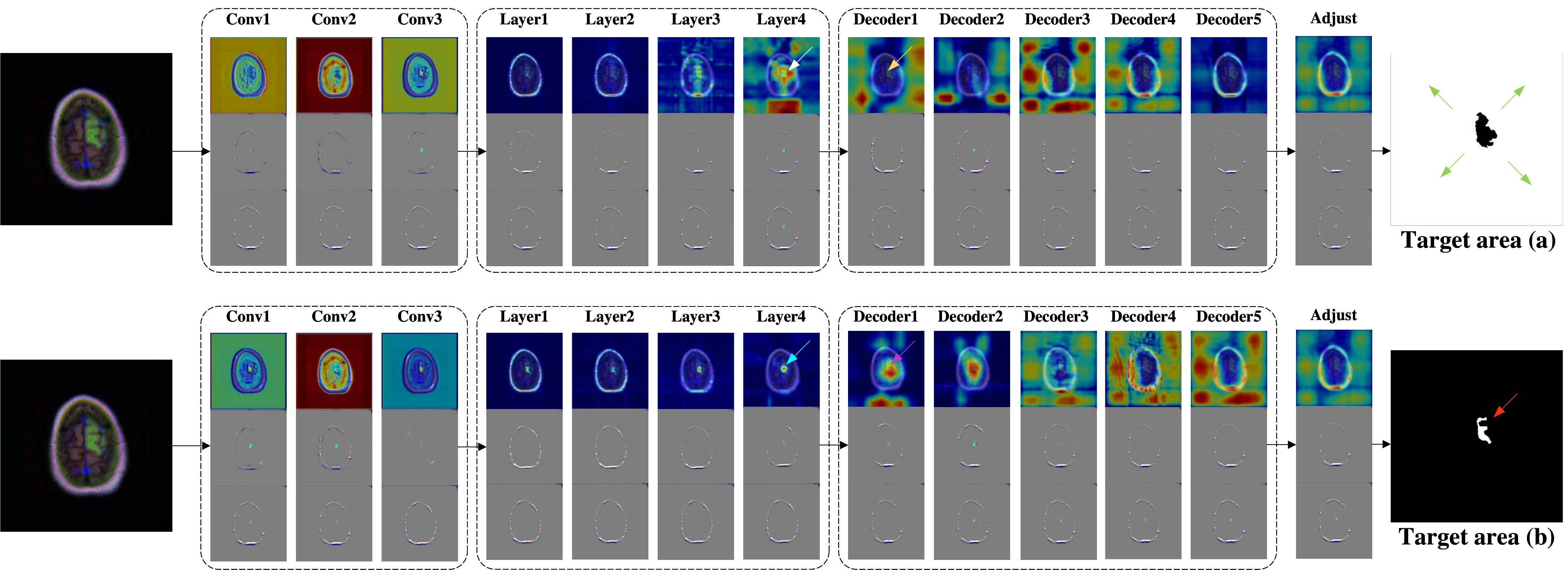}
   \caption{Visualization the internal work flows of our model based on Grad-CAM++ \cite{15Grad-CAM++}. Row-1 is Grad-CAM++ to localize class-discriminative regions; Row-2 is Guided Backpropagation \cite{17striving} to highlight all useful feature; Row-3 is Guided Grad-CAM++ to obtain high-resolution class-discrimination features for predictions. Red regions denote high prediction score and the arrows in different colors are used to highlight the important regions.}
   \label{fig:5}
   \vspace{-0.4cm}  
\end{figure*}

The neural networks were generally treated as black-box function approximations \cite{82}, leading to confusing clinical results that the medical experts cannot understand the decision-making process properly. To this end, we attempted to investigate the working flows of the proposed method by inspecting and visualizing the feature maps of the internal layers based on Grad-CAM++ \cite{15Grad-CAM++}, which can improve the decision support for the medical segmentation task. As illustrated in Figure \ref{fig:5} (corresponds to the architecture in Figure \ref{fig:3}), it can be observed that: our method generally locates the overall lesion region, then radiates around it and pays attention to the lesion boundary, finally obtains an accurate segmentation result. Specifically, the shallow convolution operations (see Conv1-Conv3) pay more attention to low-level features, \textit{e.g.,} edge and texture, while the high-level HAA blocks (see Layer1-Layer4) activate the target regions with a high score of the heat maps, \textit{e.g.}, lesion region (see blue arrow) and non-local region (see white arrow). In the decoder stage, the lesion region gradually protrudes in the target area (b), \textit{i.e.}, it starts highlighting the surrounding background of the center region to localize the target lesion. As indicated by the yellow arrow and purple arrow in the two maps of decoder1, it can be explicitly observed that our model can focus on an accurate background or foreground target region. Overall, the workflows of our model are similar to that of clinical experts. 


\subsection{Ablation Study} 
\begin{figure}[t]
  \centering
  \includegraphics[width=1\linewidth]{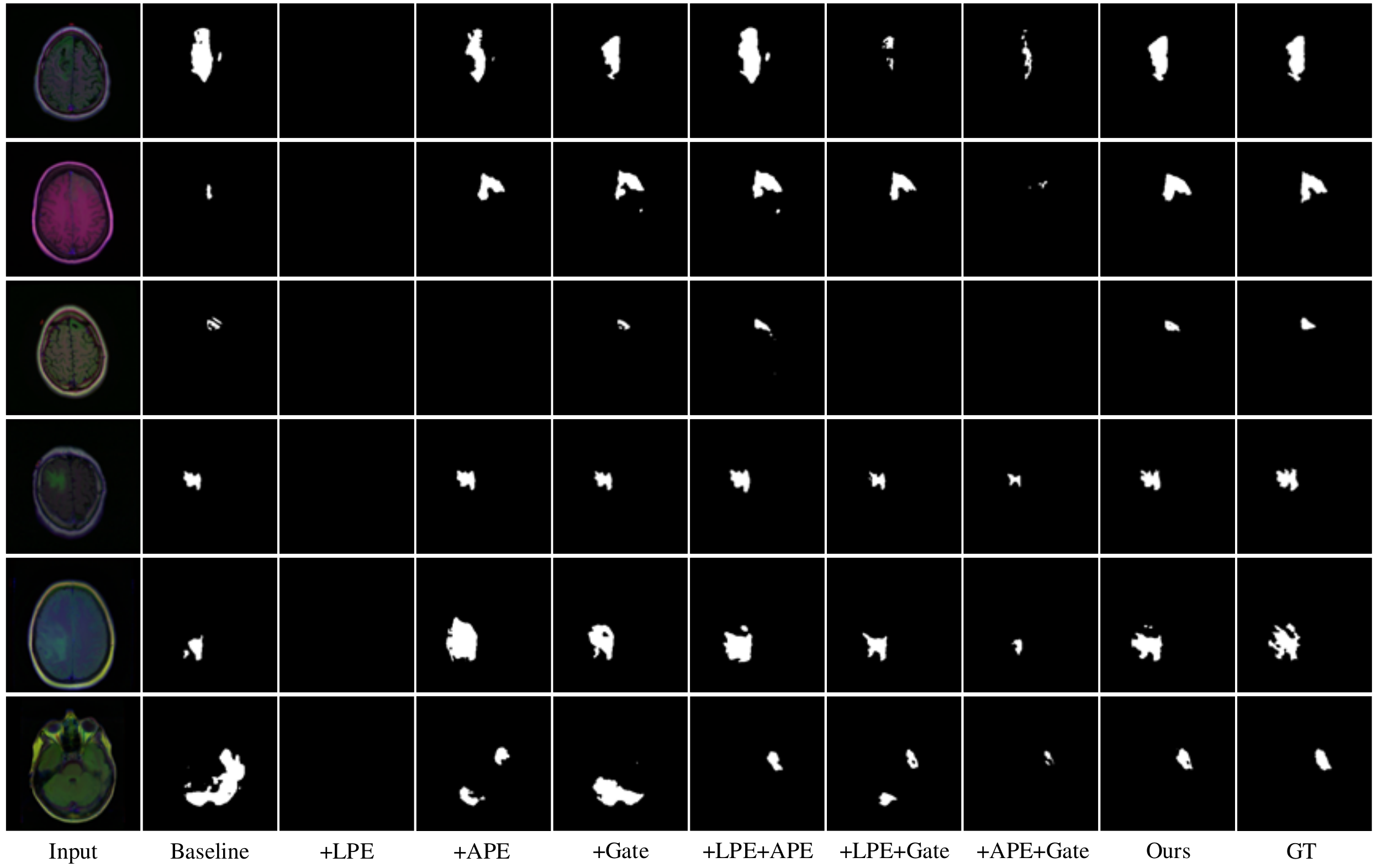}
   \caption{Visualization results of different ablations.}
   \label{fig:6}
\end{figure}

We conducted extensive ablation studies to validate the effectiveness of our key components including position embedding and gate mechanism. The quantitative and qualitative results are shown in Table \ref{tab:4} and Figure \ref{fig:6}, respectively.

\begin{table}[!t]
 \caption{Quantitative performance comparison of ablation studies.}
\setlength{\tabcolsep}{3pt}
  \centering
    \begin{tabular}{lccccccccc}
    \toprule
    \textbf{Methods} & \multicolumn{1}{c}{\textbf{AUC}} & \multicolumn{1}{c}{\textbf{MAE}} & \multicolumn{1}{c}{\textbf{WF}} & \multicolumn{1}{c}{\textbf{Dice}} & \textbf{Jacc} & \multicolumn{1}{c}{\textbf{Sensi}}\\
    \midrule
    UNet  & 80.77 & 0.91  & 74.69  & 71.40  & \multicolumn{1}{c}{60.59} & 63.67 \\
    ResUNet & 83.76 & 0.85  & 75.04 & 73.46 & \multicolumn{1}{c}{63.67} & 69.98 \\
    Baseline & 56.04 & 2.54  & 18.18 & 16.24 & \multicolumn{1}{c}{11.51} & 12.38 \\
    +LPE  & 50.29 & 2.79  & 1.17 & 1.02  & \multicolumn{1}{c}{0.62} & 0.62  \\
    +APE  & 86.70  & 1.00     & 75.27 & 75.00    & \multicolumn{1}{c}{64.54} & 76.16 \\
    +Gate & 85.86 & 0.82 & 77.51 & 76.42 & \multicolumn{1}{c}{66.41} & 74.30 \\
    +LPE+APE & \textcolor[rgb]{ .753,  0,  0}{\textbf{92.03}} & 0.92 & \textcolor[rgb]{ 0,  .69,  .314}{\textbf{78.71}} & \textcolor[rgb]{ 0,  .69,  .314}{\textbf{69.81}} & \textcolor[rgb]{ 0,  .69,  .314}{\textbf{80.2}} & \textcolor[rgb]{ .753,  0,  0}{\textbf{87.08}} \\
    +LPE+Gate & 79.53 & 1.00     & 73.47 & 69.97 & \multicolumn{1}{c}{58.39} & 61.05  \\
    +APE+Gate & 66.45 & 1.74  & 50.70  & 45.17 & \multicolumn{1}{c}{33.43} & 34.03 \\
    Ours  & \textcolor[rgb]{ 0,  .69,  .314}{\textbf{89.79}} & \textcolor[rgb]{ .753,  0,  0}{\textbf{0.70}} & \textcolor[rgb]{ .753,  0,  0}{\textbf{82.02}} & \textcolor[rgb]{ .753,  0,  0}{\textbf{81.79}} & \textcolor[rgb]{ .753,  0,  0}{\textbf{71.87}} & \textcolor[rgb]{ 0,  .69,  .314}{\textbf{82.31}} \\
    \bottomrule
    \end{tabular}%
  \label{tab:4}%
\end{table}%

\paragraph{Position Encoding.} (1) Quantitative analysis: The results in Table \ref{tab:4} clearly show that APE with fixed frequencies are necessary for boosting performance which can import prior knowledge to encode pixel-wise information, and their integration can further improve the performance especially in AUC (6.17\%) and Jaccard (22.54\%) due to the flexibility of LPE. (2) Qualitative analysis: Note that when only LPE is embeded, there is a large number of missegmented regions due to the need for large-scale datasets to learn reliable positional information. In contrast, APE tends to capture entire lesion region for its powerful spatial location information. Hence, their combination can benefit each other to capture a more accurate and complete lesion region.

\paragraph{Gate Mechanism.} (1) Quantitative analysis: Compared with the baseline, the gate mechanism improves the performance by an average of 53.28\% by focusing on the most informative feature regions. We speculated that the gate parameters can be dynamically tuned to ease the training burden, as evidenced by the fusion of LPE and Gate, with a dramatic increase (57.35\%) in performance. (2) Qualitative analysis: As shown in Figure \ref{fig:6}, LPE cannot capture any lesion regions due to a lack of data samples, but after adding gates to the model, the outline of lesion areas can be located presumably (see col-3 and col-7). Similar conclusions can be seen comparing the baseline without and with the Gate.

\paragraph{Combination of Position Encoding and Gate Mechanism.} (1) Quantitative analysis: Although their combination is slightly lower than the ensemble of APE and LPE in AUC and Sensitivity since the position embedding is capable of encoding spatial pixel-wise information, ours outperformed all the other indicators by nearly 2.26\%, which indicates the combination can encode features more efficiently on small-scale datasets. (2) Qualitative analysis: Ours is very close to the ground truth, but APE tends to over-capture the lesion region (see col-4), while only using Gate tends to ignore the contour of the lesion region. 



\section{Conclusions}
In this work, we proposed a parameter-efficient Transformer with a novel HAA for medical image segmentation, which can embed more intrinsic inductive biases into learning. In particular, we empirically investigated how different position encoding strategies affect the prediction quality of ROI and demonstrated that APE is necessary for pixel-level image segmentation tasks, which outperforms RPE in missed detection rate when foreground and background are highly similar. In addition, we explored more inductive biases into the network via regularization technique, \textit{i.e.}, the gate mechanism. In future work, we plan to combine axial-attention mechanisms with direction-aware and channel-sensitive information, and exploit the impact of orientation on ROI in identifying visual structural information.

\bibliographystyle{unsrt}  
\bibliography{egbib}  

\end{document}